# Cationic ordering control of magnetization in $Sr_2FeMoO_6$ double perovskite

Ll. Balcells, J. Navarro, M. Bibes, A. Roig

B. Martínez, and J. Fontcuberta[§]

Institut de Ciència de Materials de Barcelona (CSIC).

Campus Universitat Autònoma de Barcelona, E-08193 Bellaterra. Catalunya.

Spain

**Abstract**

The role of the synthesis conditions on the cationic Fe/Mo ordering in $Sr_2FeMoO_6$ double perovskite is addressed. It is shown that this ordering can be controlled and varied systematically. The Fe/Mo ordering has a profound impact on the saturation magnetization of the material. Using the appropriate synthesis protocol a record value of $3.7\mu_B$/f.u. has been obtained. Mössbauer analysis reveals the existence of two distinguishable Fe sites in agreement with the P4/mmm symmetry and a charge density at the $Fe^{m+}$ ions significantly larger than $3d^5$ suggesting a Fe contribution to the spin-down conduction band. The implications of these findings for the synthesis of $Sr_2FeMoO_6$ having optimal magnetoresistance response are discussed.

Although oxides of the type $A_2BB'O_6$ where A is an alkaline earth (A=Sr, Ca, Ba) and B, B' are heterovalent transition metals such as B=Fe, Cr, .. and B'= Mo, W, Re,... , have known since long ago [1,2] they are receiving a renewed great deal of attention. This is motivated by the recent report that $Sr_2FeMoO_6$ is a half-metallic ferromagnet with a relatively high Curie temperature (about 410-450K) [3]. Its half metallic nature leads to an ideal full polarization of the itinerant carriers and thus these materials are viewed as a serious alternative to the much investigated manganese perovskites but with the added advantage of having a wider temperature range for practical applications as magnetoresistive materials.

The structure is built up by ordering perovskite blocks in a rock salt superlattice and the properties of the material are thought to critically depend on this ordering. $Sr_2FeMoO_6$ is believed to be ferrimagnetic-like, i.e. the B and B' sublattice are antiferromagnetically coupled. In the simplest ionic picture $Fe^{3+}$($3d^5$, S=5/2) ions in B sites are antiferromagnetically coupled to its six $Mo^{5+}$($4d^1$, S=1/2) neighbors occupying the B' sites and thus a saturation magnetization $M_S = 4\mu_B$ is predicted. Accordingly, it is expected that $M_S$ should be sensitively dependent on the ordering of Fe/Mo ions among the B/B' sublattices. Indeed, the $M_S$ values reported so far are systematically much smaller ($3.1\mu_B$ [3], $3.5\mu_B$ [4] $3.2\mu_B$ [5]), than the predicted $4\mu_B$ value. It is commonly thought that this significant decrease is due to antisite defects resulting from the partial disorder of Fe and Mo ions among the B/B' sublattices. Montecarlo simulations have indeed predicted a reduction of $M_S$ as a function of the antisite disorder that could account for the experimental observations [6].

However, there are no strong experimental evidences that Fe/Mo disorder is the reason for the observed reduction of $M_S$ and thus strategies to enhance $M_S$ are lacking. We also note that the simple $Fe^{3+}/Mo^{5+}$ ionic picture needs to be validated as neutron diffraction experiments have failed to observe significant moment at the Mo sites [7].

In this letter we will show that the concentration of antisite defects (AS≡Concentration of Fe ions at the Mo sites/f.u.) can be appropriately controlled by the synthesis protocol and that by doing so $M_S$ can be varied monotonically. The observed variation of $M_S$ on AS provide a unique experimental confirmation of the ferrimagnetic ordering in the lattice. Using this approach we have obtained a record $M_S \approx 3.7$ $\mu_B$ value.

Ceramic $Sr_2FeMoO_6$ samples were prepared by solid state reaction. Appropriate mixtures of $CO_3Sr$, $Fe_2O_3$ and $MoO_3$ were intimately mixed in an agate mortar. Calcination was performed at 900ºC for 4h under air. A subsequent high temperature reaction was performed at various temperatures $T_S$ (900ºC, 950ºC, 1000ºC, 1100ºC, 1200ºC) under 5%$H_2$/Ar stream for 16h. Samples were cooled to room-temperature at 300ºC/h. X-ray diffraction powder patterns were collected using Cu-k$_\alpha$ radiation using a Rigaku Ru-200B diffractometer. Inspection of the patterns revealed that almost pure single tetragonal P4/mmm symmetry phase is obtained. The most relevant trend observed in the recorded patterns is the systematic increase of the diffracted intensity ratio I(101)/{I(200)+I(112)}. In Fig. 1 we show portions of the θ-2θ patterns displaying the angular region of interest. In Fig. 2 the ratio I(101)/{I(200)+I(112)} is depicted as a function of $T_S$. We recall here that (101) is a superstructure line which reflects the ordering of Fe/Mo ions at B/B' sites. Therefore, the results of Figs. 1 and 2 clearly indicate that cationic ordering is progressively achieved when rising the sintering temperature $T_S$.

Detailed quantitative Rietveld analysis of the patterns has also been used. In the P4/mmm space group, when the Fe/Mo ions are fully ordered, the Fe ions occupy two non-equivalent positions (1a and 1d) whereas Mo ions occupy the corresponding 1b and 1c positions. Antisite defects concentration AS can be determined by allowing the occupancy of Fe and Mo ions in both sets of positions to be varied. In Fig. 2 we show the obtained AS values for each sample as a function of $T_S$. In agreement with the trend already observed from the $I(101)/\{I(200)+I(112)\}$ variation, cationic ordering is improved by high temperature sintering. Indeed AS$\approx$2% is obtained for $T_S$= 1200ºC. We also point out that for the sample prepared at the lowest $T_S$, minor traces of $SrMoO_4$ have been identified but its concentration does not exceed the 1% (from Rietveld analysis) and its is absent for samples sintered at higher temperatures.

The magnetization of the samples has been measured by using a commercial SQUID magnetometer. In Fig. 3a we show the magnetization *vs* field curves M(H) recorded at 10K for samples sintered at given temperatures. Inspection of these data immediately reveals that the saturation magnetization of the samples systematically lowers when lowering $T_S$. From the above discussion it follows that when increasing the disorder the magnetization indeed decreases. This can be better appreciated in Fig. 3 where we plot $M_S$ *vs* AS as obtained from the Rietveld refinement for each sample. It is worth noting that our $T_S$=1200ºC sample, having a disorder of only 2%, has a saturation of 3.7$\mu_B$, which is the highest so far reported for $Sr_2FeMoO_6$. Data in Fig. 3b reveals that $M_S$ has an almost linear dependence on the antisite concentration. In fact, a linear regression ($M_S$=a+b*AS) of the experimental data (solid line in Fig. 3b) leads to a= 4.04(9)$\mu_B$ and b=8.1(8)$\mu_B$/f.u.. The significance of this fit will be fully appreciated if the expected dependence of $M_S$ on the antisite concentration is evaluated for the simplest ferrimagnetic arrangement. Lets call $m_B$ and $m_{B'}$ the magnetizations of the B and B' sublattices. Therefore an antiferromagnetic coupling between B and B' will produce a net magnetization of $M_S$= $m_B$ - $m_{B'}$. If sublattice B is occupied by (1-x) Fe ions and x Mo ions, then $M_S$=(1-2x)$m_{Fe}$-(1-2x)$m_{Mo}$ where $m_{Fe}$ and $m_{Mo}$ are the magnetic moment of the Fe and Mo ions respectively and x≡AS. This simple picture, assuming a spin-only contribution, $m_{Fe}(Fe^{3+})=5\mu_B$ and $m_{Mo}(Mo^{5+})=1\mu_B$, leads to $M_S=(4-8x)\mu_B$ (dashed line in Fig. 3b). The Montecarlo computation of Ogale et al [6] also predicted $M_S=(4-8x)\mu_B$. Alternatively, if one assumes that locally, at the antisites, the superexchange rules for $3d^5$-$3d^5$, $4d^1$-$4d^1$ and $4d^1$-$3d^5$ holds, then $M_S=(4-10x)\mu_B$ is predicted (doted line in Fig. 3b). The agreement between the predictions of these simple models and the experimental data of Fig. 3b is remarkable.

It should be noticed that previous analysis of the dependence of $M_S$ on the antisite concentration is still valid even if the effective charge state of the distinct species were: $Fe^{(3-\delta)+}$ and $Mo^{(5+\delta)+}$, i.e $m_{Fe}(Fe^{(3-\delta)+})=(5-\delta)\mu_B$ and $m_{Mo}(Mo^{(5+\delta)+})=(1-\delta)\mu_B$, Therefore, to assess the charge density on each ion additional information is required. In Fig. 4 we show the room-temperature Mössbauer spectrum of the $T_S=1200°C$ sample, recorded using a $^{57}$CoRh source. Inspection of this spectrum reveals the existence of two hyperfine splitted sextets (I,II), characteristic of magnetic ordering and having similar resonant areas ($A_I$ and $A_{II}$). Fitting of the spectrum using two sextets, having the conventional 3:2:1:1:2:3 intensity relationship allows extracting the Isomer Shift (IS) values $IS_I \approx IS_{II} \approx 0.65(6)$mm/s (relative to α-Fe) and hyperfine fields (HF) of 330(1)kOe and 294(1)kOe for I and II respectively, having intensities of $A_I=57(2)\%$ and $A_{II}\approx 43(6)\%$ respectively. We first notice that two sites for Fe ions, having almost identical coordination, are expected from the P/4mmm space group symmetry. Therefore the Mössbauer data is fully consistent with the structural analysis. The departure of $A_I:A_{II}$ from its ideal ratio (1:1) may indicate of the existence of some residual Fe/Mo disorder in this sample. Probably of more relevance is the observation that the hyperfine parameters (IS and HF) obtained are much different from those typically found for $Fe^{3+}$ ions in an octahedral oxygen environment. Extensive data analysis will be reported separately [8]. Here we should mention that the IS and HF parameters are both a measure of the charge density of the ion ($Fe^{m+}$). Indeed the experimental values of IS and HF indicate that m≈2.6 thus corresponding to a formal electronic configuration of $3d^{5+\delta}$ with $\delta= 0.4$ [9].

Kobayashi et al [3] have shown that the spin up subband of Fe ions is full for a $3d^5$ configuration. Therefore, the observed excess electrons ($\delta$) should go into the minority spin-down subband ($t_{2g}^{\downarrow}$). According to calculations, in this oxide the conduction band is formed by hybridized Fe-$t_{2g}^{\downarrow}$ and Mo-$t_{2g}^{\downarrow}$ orbitals [3]. Our experimental Mössbauer data is in excellent agreement with this picture. It also indicates that the localized ionic picture used to evaluate the $M_S(AS)$ dependence may not be accurate enough and a more refined model will be required. Even though, the agreement between the experimental and calculated $M_S$ values indicates that it collects the main contribution to the saturated magnetization. For samples synthesized at lower $T_S$ the room temperature Mössbauer spectra become gradually more complex due to the existence of multiple coordination for the Fe-probe nuclei.

In summary, we have shown that suitable synthesis conditions allow obtaining $Sr_2FeMoO_6$ samples with variable degree of Fe/Mo cationic ordering among the B and B' sublattices. This ordering translates into the saturation magnetization and thus to the effective spin polarization of the oxide which can be controlled and optimized. Its is thought that the optimal polarization (100%) is required to take full advantage of the intrinsic half-metallic nature of these oxides for tunnel-magnetoresistance based devices. We have also provided strong evidence of the contribution of Fe-3d electrons to the conduction itinerant band.

After submission of the present paper, Yin et al [10] suggested that antiphase boundaries may play a role on the reduction of magnetization. Although the presence of antiphase boundaries in our samples can not excluded, the reduction of the superstructure diffraction peak (Figs. 1 and 2) when can not be explained by antiphase boundary defects but rather by antisite point defects.

Financial support by the CICYT (MAT99-0984), and the CEE-OXSEN projects and the Generalitat de Catalunya (GRQ99-8029) are acknowledged.

**Figure Captions**

**Fig. 1** Portion of the diffracted intensity (log. scale) $\theta$-$2\theta$ patterns displaying the (101) - superstructure- and the (200) and (112) reflections for $Sr_2FeMoO_6$ samples prepared at various temperatures $T_S$.

**Fig. 2** Relative intensity $I(101)/\{I(200)+I(112)\}$ (right axis) and antisite concentration (AS) (left axis) as a function of $T_S$.

**Fig. 3 a)** Magnetization curves at 10K measured for a number of samples prepared at various temperatures $T_S$. **b)** Dependence of the saturation magnetization on the antisite concentration AS. Solid line is the lineal fit of the experimental data; Dashed (labeled FIM) and Dotted (labeled SE) lines represent the expected dependence of $M_S(AS)$ within the two models discussed in text.

**Fig. 4.** Room-temperature Mössbauer spectrum of the $T_S=1200°C$ $Sr_2FeMoO_6$ sample.

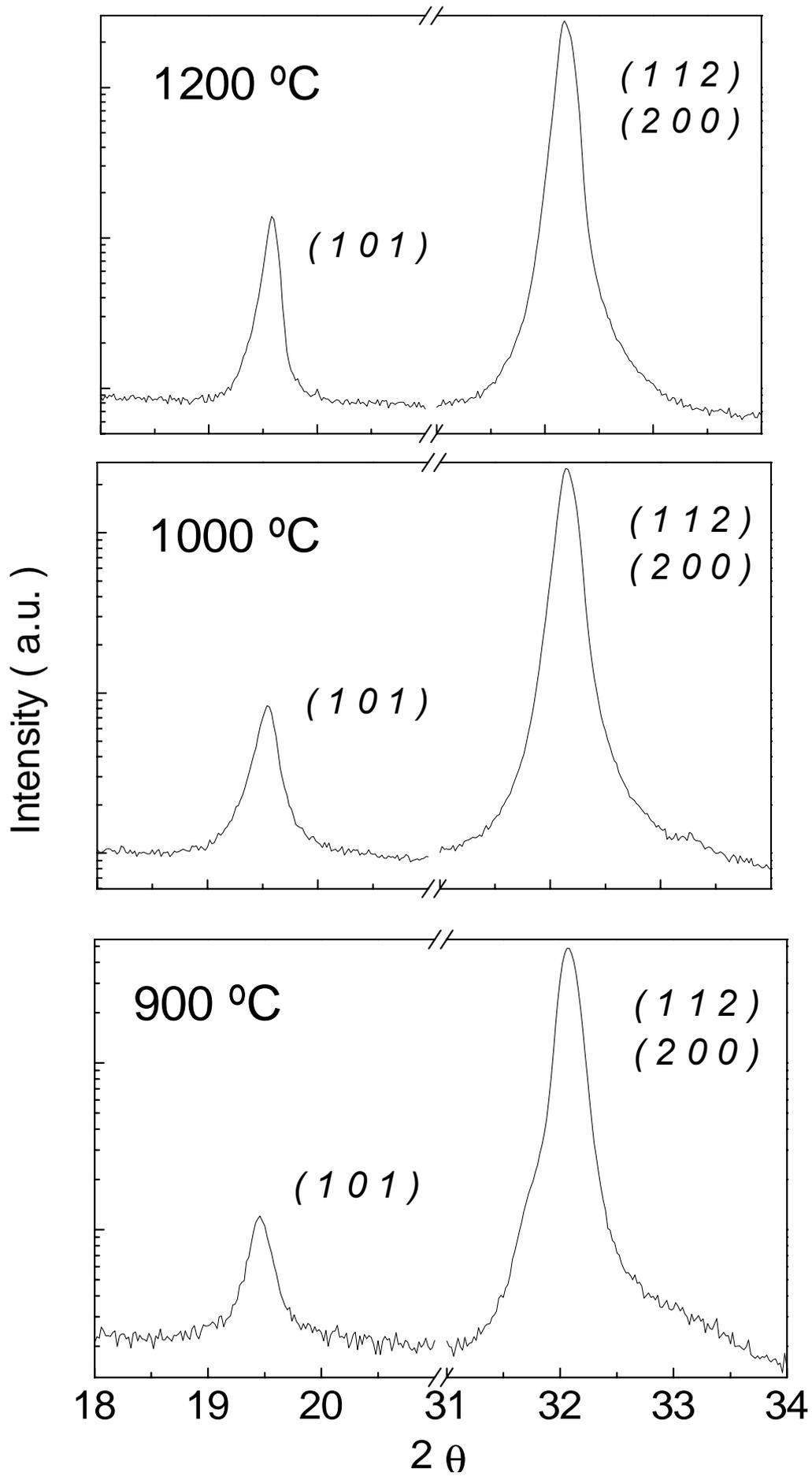

Fig.1

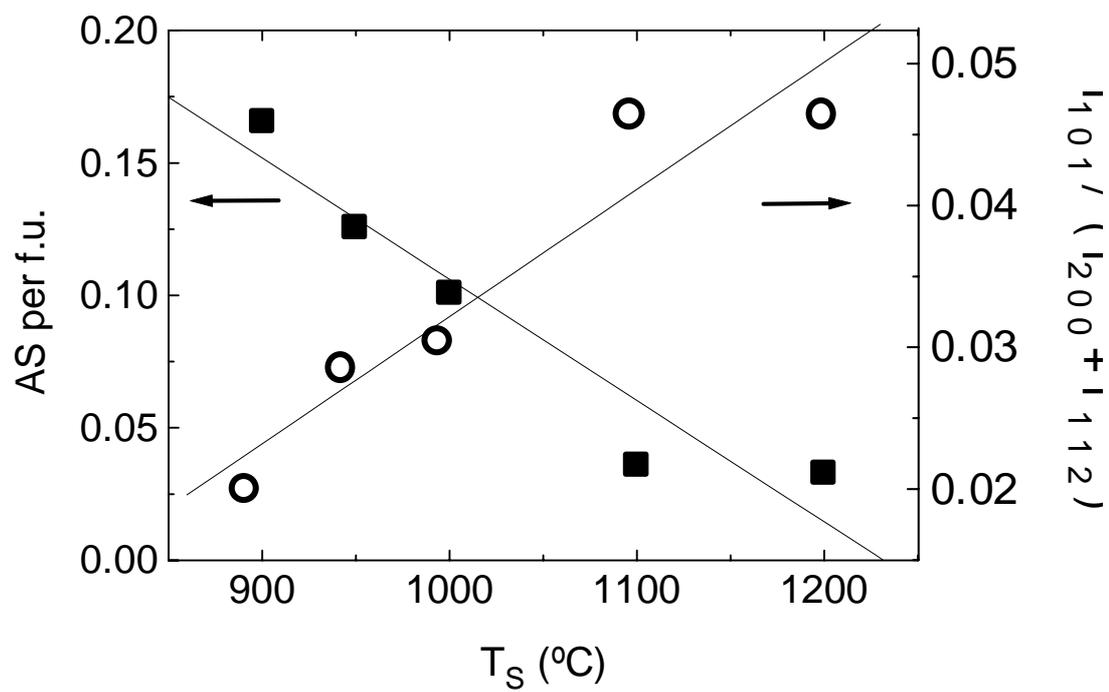

**Fig. 2**

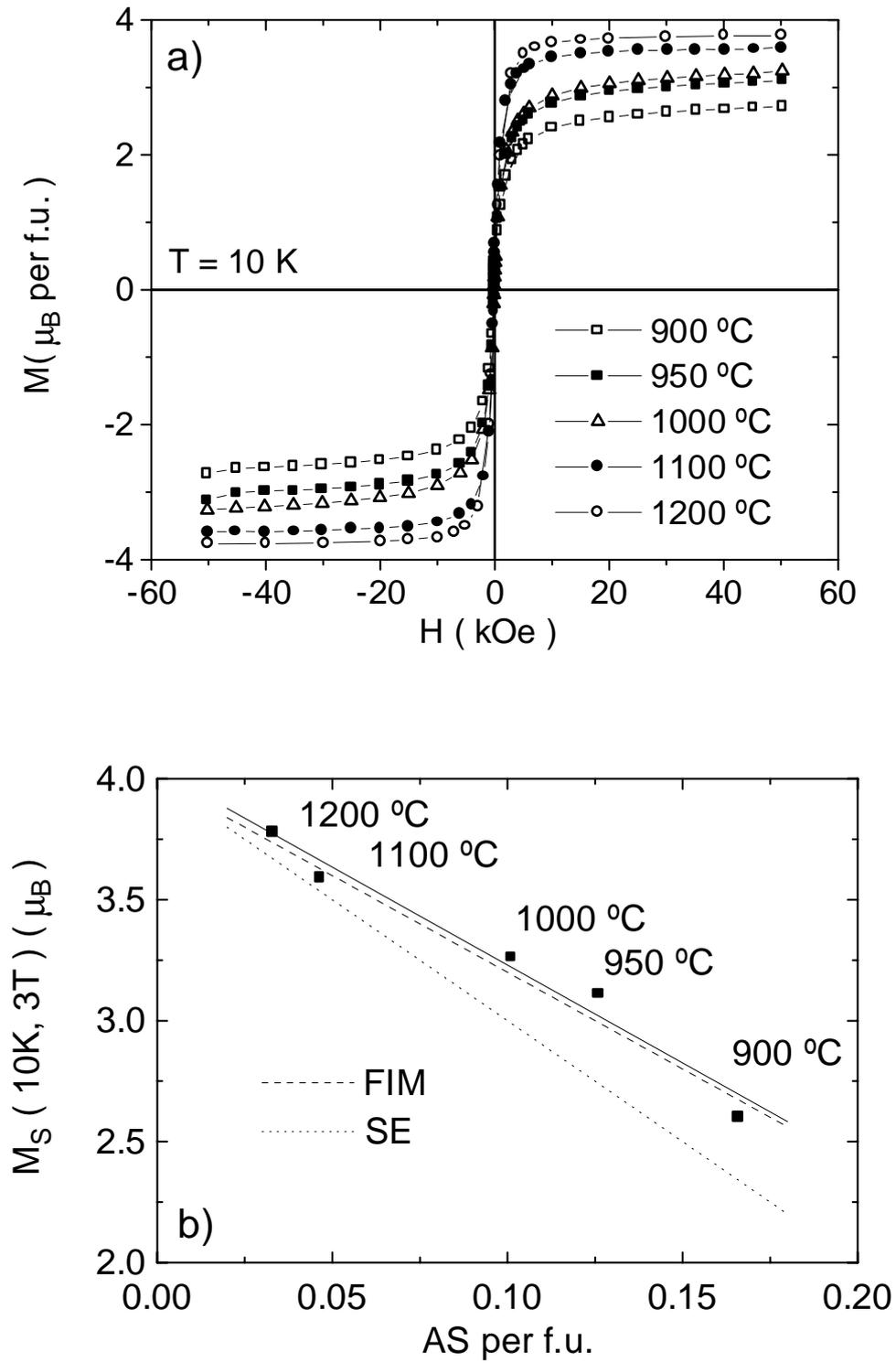

Fig. 3

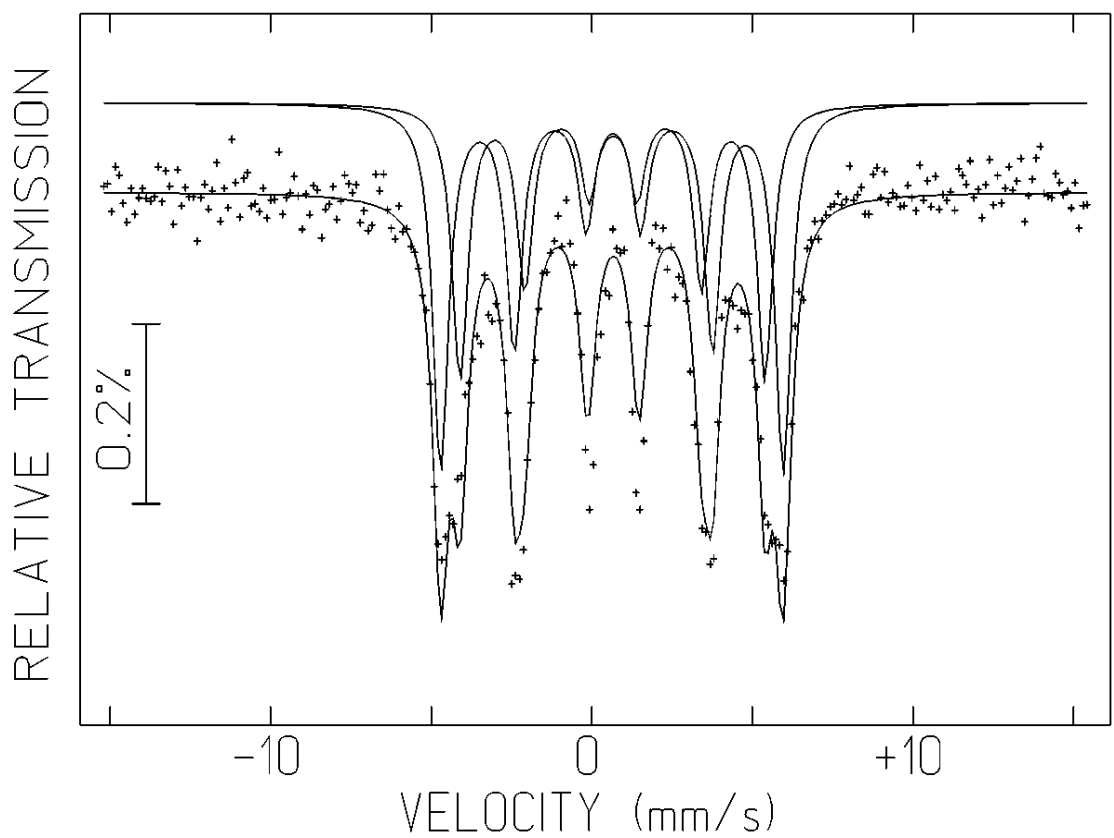

Fig. 4